\documentclass[%
reprint,
 amsthm,amsmath,amssymb,
 aps,mathrsfs,
]{revtex4-2}

\usepackage{graphics}
\usepackage{graphicx}
\usepackage{amsmath}
\usepackage{longtable}
\usepackage{url}
\usepackage[caption=false]{subfig}
\usepackage{graphicx,bm}
\usepackage{verbatim}
\usepackage{color}
\usepackage{lipsum}
\usepackage{mathtools, nccmath}

\begin{document}

%\preprint{APS/123-QED}

\title{%Angular momentum dependence of three fermions in the unitarity limit
Efimov physics implications at $p$-wave fermionic unitarity}% Force line breaks with \\

\author{Yu-Hsin Chen}
\email{chen2662@purdue.edu}
\affiliation{Department of Physics and Astronomy, Purdue University, West Lafayette, Indiana 47907 USA}
\author{Chris H. Greene}
\email{chgreene@purdue.edu}
\affiliation{Department of Physics and Astronomy, Purdue University, West Lafayette, Indiana 47907 USA}
\affiliation{Purdue Quantum Science and Engineering Institute, Purdue University, West Lafayette, Indiana 47907 USA}


\date{\today}% It is always \today, today,

\begin{abstract}
Efimov physics at $p$-wave unitarity for three equal mass fermions in multiple symmetries interacting via Lennard-Jones potentials is predicted to modify the long range interaction potential energy, but without producing a true Efimov effect. This analysis treats the following total orbital angular momenta and parities, $J^{\Pi}=0^{+}, 1^{+}, 1^{-}$ and $2^{-}$, for either three spin-polarized fermions ($\uparrow \uparrow \uparrow $), or two spin-up and one spin-down fermion ($\downarrow \uparrow \uparrow $). Our results for the long range interaction in some of those cases agree with previous work by Werner and Castin and by Blume {\it et al.}, namely in cases where the $s$-wave scattering length goes to infinity. The present results extend those calculated interaction energies to small and intermediate hyperradii comparable to the van der Waals length, and we consider additional unitarity scenarios where the $p$-wave scattering volume approaches infinity.  The crucial role of the diagonal hyperradial adiabatic correction term is identified and characterized. %Our study also considers different interactions between the atoms in different spin states, such as the case where the two spin-up fermions have a $p$-wave interaction and where a spin-up atom interacting with a spin-down atom has a strong $s$-wave interaction as well. Another case treated involves the different spin state fermions with strong $p$-wave interaction, and the two identical spin state fermions with weak or strong $p$-wave interaction. We further consider three spin-up fermions at the $p$-wave unitarity limit with different symmetries. Universal three-body channel properties are derived for the above cases. Another finding is that there are one or two new $p$-wave unitary channels for different total orbital angular momenta $J^{\Pi}$, which also show universal behavior.  Moreover, our results demonstrate that the presence of $p$-wave unitary two-body interaction(s) modifies some of the lowest three-body continuum long-range potentials. 
%\begin{description}
%%\item[Structure]
%\end{description}
\end{abstract}

\maketitle

%\tableofcontents

%\section{\label{sec:level1}Background and Introduction %\protect\\ The line
%break was forced \lowercase{via} \textbackslash\textbackslash}

\maketitle

\section{Introduction}
In recent decades it has become routine to manipulate interactions in the dilute Fermi gas using a Feshbach resonance to produce a quantum system at or near the unitary limit \cite{chin2010feshbach,greiner2003emergence,regal2004observation,zwierlein2003observation,greene2017universal}. For instance, the $s$-wave scattering length between fermions in different internal quantum states diverges to infinity $a_{s}=\pm\infty$, entering a universality regime where the specific short-range details about the underlying potential energy function are to a large extent irrelevant \cite{werner2006unitary}. Many $a_{s}\rightarrow \infty$ physical  phenomena in a two-component Fermi gas are by now well known and extensively studied; this regime is often denoted the BCS-BEC crossover problem \cite{blume2007universal,von2007spectrum}. When $a_{s}>0$ is large, two fermions in different spin states can pair to form a bosonic molecule, while for $a_{s}<0$ two different spin fermions can form a Cooper pair \cite{strecker2003conversion,bourdel2004experimental,petrov2004weakly}. In the unitary regime, the physics is rich, nonperturbative, and challenging. Our main interest in the present study is in generalizing our understanding of such systems having a very large $p$-wave scattering volume.  This interest has been sparked in part by recent experiments relating to the three-body loss rate and the unitary limit of spin-polarized Fermi gases \cite{yoshida2018scaling,waseem2018unitarity,waseem2019quantitative}, as well as theoretical investigations that have predicted and analyzed the mechanisms of recombination or loss in a single-component fermion system~\cite{suno2003recombination,suno2003three,jona2008three,schmidt2020three}. There are also now multiple ways of controlling ultracold atom-atom interactions, via not only magnetic Fano-Feshbach resonances, but also light-induced or rf-induced resonances or orbital resonances;  these experimental tools can in principle be used to independently modify both $s$-wave and $p$-wave interactions in a few-body system\cite{fedichev1996influence,theis2004tuning,tscherbul2010rf}.

A major conjecture of this study is that all $N$-body systems having finite range interactions, with at least some of their interactions equal to their unitary limit, will generally show a key consequence of Efimov physics at unitarity: namely, a reduction of some of the asymptotic adiabatic channel potentials (including normally the lowest) in their coefficient of $1/R^2$, where $R$ is the hyperradius.  This reduction appears to be a universal effect that occurs at unitarity, and has previously been demonstrated at $s$-wave unitarity for many 3-body systems and for some 4-body systems. (See, for instance, recent evidence presented for 3- and 4-nucleon systems with their $s$-wave interactions tuned to unitarity \cite{Higgins1,Higgins2}).  Only in some cases, such as three equal mass bosons with all pairwise scattering lengths infinite, does this asymptotic coefficient become reduced all the way to negative values, in which case there is a true {\it Efimov effect} with the usual infinite sequence of weakly-bound energy levels converging geometrically to zero energy.  In the cases involving $p$-wave interactions at unitarity that are considered in the present study, we document the role of {\it Efimov physics} in the aforementioned sense, but there is no true {\it Efimov effect} for any of the symmetries treated. 

The adiabatic hyperspherical representation serves as our main theoretical tool to analyze these different flavors of the three-fermion problem. The present focus is mainly on the system with two spin-up and one spin-down system in various symmetries $J^{\Pi}$ \cite{esry2001threshold} having total orbital angular momentum $J$ and parity $\Pi$. For this case, with no $p$-wave interactions, we have obtained accurate numerical results for the symmetries $J^{\Pi}=0^{+}$ and $1^{-}$. Moreover, our work goes beyond $s$-wave unitarity studies, by introducing the $p$-wave scattering volume characterizing the interaction between two spin-up fermions that is varied to reach very large magnitudes approaching infinity.  Remarkably, the adiabatic 3-body potential curves are sometimes modified by the presence of $p-$wave unitarity in one or more of the pairwise interactions. Our results show the emergence of a novel class of three-body fragmentation channels in the $p$-wave unitary limit, which differ for different angular momentum and parity symmetry classes.

Different variations of adiabatic approximation or of the Born-Oppenheimer approximation arise and are utilized in the physical chemistry of diatomic molecules, where the coordinate treated adiabatically is the internuclear distance.\cite{VanVleck1936jcp,JungenAtabek1977}  
In the present study where the adiabatic coordinate is the hyperradius $R$, it is important to note that similar issues arise.  Specifically, the starting point of the adiabatic hyperspherical representation is solution of the Schr\"odinger equation with $R$ held fixed, which yields eigenvalues $U_\nu(R)$ that we denote the {\it Born-Oppenheimer} potential curves.  When the hyperradial Schr\"odinger equation is solved using only these $U_\nu(R)$ for bound states or scattering or resonance states, in a single-channel approximation, these results are correspondingly denoted as coming from the Born-Oppenheimer approximation. When the diagonal term $Q_{\nu\nu}(R) $ proportional to the second derivative in $R$ is included in the potential curve, this is denoted the {\it adiabatic approximation} and it is nearly always more accurate than the Born-Oppenheimer approximation.  This terminology is important for calculations based on the adiabatic hyperspherical representation because examples have been found (to be further documented below in the present study) where the Born-Oppenheimer potential curves (i.e. that neglect the diagonal elements of $Q$) would appear to suggest that the Efimov effect\cite{Efimov1971} occurs in that symmetry, but after $Q_{\nu\nu}(R) $ is included, the apparent Efimov effect goes away.  Moreover, coupled channels calculations confirm in those cases that there is no true Efimov effect.  Parenthetically, it should be noted that in the standard Efimov scenario for 3 equal mass bosons at unitarity, $Q_{\nu\nu}(R) \rightarrow 0 $, so the Efimov potential already arises for the Born-Oppenheimer long-range potential, and has no difference from the adiabatic potential curve, at least for zero-range interactions.

The possibility of an $s$-wave Efimov effect in equal mass two-component three-fermion systems have been extensively investigated over the past 15-20 years\cite{petrov2003three,blume2010breakdown,liu2010three,peng2011high}. The existence of a $p$-wave Efimov effect in the one-component Fermi gas was initially predicted to occur by Macek \textit{et al}. \cite{macek2006properties} and by Braaten \textit{et al}. \cite{braaten2012renormalization} for the two-component Fermi gas, before their arguments were shown to be incorrect by Nishida \cite{nishida2012impossibility}. 

The Efimov effect\cite{Efimov1971} is defined here as occurring whenever there is a long-range hyperspherical potential curve for a trimer system of the following form,
\begin{equation}
W(R) \rightarrow \frac{s^2-\frac{1}{4}}{2 \mu R^2},
\end{equation}
where $\mu=m/\sqrt{3}$ is the hyperspherical 3-body reduced mass and where $s^2<0$.
Our own study here confirms that the $p$-wave Efimov effect, which if it exists would imply that the system possesses an infinite series of energy levels converging geometrically to zero energy, does not occur for any system of three equal mass fermions with at least two in identical spin states, regardless of the $s-$ or $p$-wave interaction strength.    However, the Efimov-like modification of the long-range 3-body continuum potentials does result in modified threshold laws at unitarity for the symmetries considered.

\section{Method}
The present study considers two different classes of fermionic systems: either three spin-polarized fermions ($\uparrow \uparrow \uparrow $)  or two spin-up fermions interacting with one spin-down fermion ($\downarrow \uparrow \uparrow $) with total angular momenta $J^{\Pi}=0^{+}, 1^{+}, 1^{-}$ and $2^{-}$, where $\Pi$ is the total parity in the system. First, the Schr\"{o}dinger equation is rewritten using  modified Smith-Whitten hyperspherical  coordinates~\cite{whitten1968symmetric,suno2008adiabatic,suno2002three}:
\begin{align}\label{eq:Schrodingereq}
	\left[-\frac{1}{2\mu}\frac{\partial^{2}}{\partial R^{2}}+\frac{15\hslash^{2}}{8\mu R^{2}}+\frac{\Lambda^{2}}{2\mu R^{2}}+V(R,\theta,\phi)\right]\psi_{E}=E\psi_{E}.
\end{align}
Here $\Lambda^{2}$ is the squared "grand angular momentum operator" and $\mu = m/\sqrt{3}$ is the three-body reduced mass for three identical fermions with mass $m$. The interaction potential $V(R,\theta,\phi)$ is taken to be a sum of the two-body potentials
\begin{equation}
    V(R,\theta,\phi) = v_3(r_{12})+v_1(r_{23})+v_2(r_{31})
\end{equation}
where the $r_{ij}$ are the interparticle distance.  The two-body potential adopted thus far is the Lennard-Jones potential~\cite{wang2012origin}
\begin{equation}
	v_n(r)=-\frac{C_{6}}{r^{6}}\left(1-\frac{\lambda_n^{6}}{r^{6}}\right)
\end{equation}
In the present treatment the parameter $\lambda_n$ is used to adjust the values to get desired $s$-wave scattering length or $p$-wave scattering volume for a chosen pair of atoms. The two-body $p$-wave scattering volume can be represented as~\cite{suno2003recombination} 
\begin{equation}
	V_{p}=-\lim_{k \to 0}\left(\frac{\tan\delta_{1}(k)}{k^{3}}\right)
\end{equation}
where the $\delta_{1}(k)$ is the $p$-wave scattering phase shift and $k$ is the wave number. 
To solve the Schr\"{o}dinger equation Eq.(\ref{eq:Schrodingereq}), we use the adiabatic representation.  This requires first fixing the $R$ value, neglecting all $R$ derivatives, and solving the adiabatic eigenvalue equation in the hyperangles for the chosen symmetry $J^{\Pi}$, where $J$ represents the total orbital angular momentum and $\Pi$ is the total parity:
\begin{equation}\label{eq:adiabaticeq}
    \left[\frac{\Lambda^{2}}{2\mu R^{2}}+\frac{15\hslash^{2}}{8\mu R^{2}}+V(R,\theta,\phi)\right]\Phi_{\nu}(R;\Omega)=U_{\nu}(R)\Phi_{\nu}(R;\Omega).
\end{equation}
Here we denote the eigenvalues $U_{\nu}(R)$ as the {\it Born-Oppenheimer potential curves} and the eigenfunctions $\Phi_{\nu}(R;\Omega)$ are the corresponding channel functions. $\Omega$ represents the five hyperangles $\Omega \equiv (\theta, \varphi, \alpha, \beta, \gamma)$ plus any relevant spin degrees of freedom. The adiabatic representation then expands the full desired wave function $\psi_{E}(R;\Omega)$ in a truncated subset of the complete orthonormal set of hyperangular eigenfunctions $\Phi_{\nu}(R;\Omega)$, each multiplied by a corresponding radial wave function $F_{\nu E}(R)$ to be determined \cite{suno2009three},
\begin{equation}\label{eq:totalwavefunc}
	\psi_{E}(R;\Omega)=\sum_{\nu=0}^{\infty}F_{\nu E}(R)\Phi_{\nu}(R;\Omega).
\end{equation}
In order to diagonalize the adiabatic Hamiltonian Eq.(\ref{eq:adiabaticeq}), our treatment follows the standard route that expands the channel function into Wigner D functions which rotate from the laboratory frame into a body-frame coordinate system:
\begin{equation}
	\Phi_{\nu}(R;\Omega)=\sum_{K}^{J}\phi_{K\nu}(R;\theta,\varphi)D_{KM}^{J}(\alpha, \beta, \gamma).
\end{equation}
The quantum numbers $K$ and $M$ represent the projection of $\vec{J}$ onto the body-fixed and space-fixed $z$-axes, respectively. K takes the values $J,J-2,...,-(J-2),-J$ for the "party favored" case and $J-1, J-3,...,-(J-3),-(J-1)$ for the "party unfavored". For our cases $J=0^{+}, 1^{+}, 1^{-}$ or $2^{-}$; setting $M=0$, the channel function can be written as 
\footnotesize
\begin{align}
    &\Phi_{\nu}(R;\Omega)=\phi_{\nu}(R;\theta,\varphi) \quad \left(J=0^{+}\right) \\
    &\Phi_{\nu}(R;\Omega)=\phi_{\nu}(R;\theta,\varphi)\cos\beta \quad  \left(J=1^{+}\right) \\
	&\Phi_{\nu}(R;\Omega)=\phi_{\nu c}(R;\theta,\varphi)\sin\beta \cos \gamma  \nonumber\\
	&\qquad\qquad + i \phi_{\nu s}(R;\theta,\varphi)\sin\beta \sin \gamma  \quad \left(J=1^{-}\right) \\
	&\Phi_{\nu}(R;\Omega)=\phi_{\nu c}(R;\theta,\varphi)\sin (2\beta) \cos \gamma \nonumber\\
	&\qquad\qquad+ i \phi_{\nu s}(R;\theta,\varphi)\sin (2\beta) \sin \gamma  \quad \left(J=2^{-}\right) 
\end{align}
\normalsize
Particles 2 and 3 are designated as the ones occupying the same spin state, whereby the permutation operation $P_{23}$ corresponds to\cite{suno2009three}
\begin{equation}
	\varphi \rightarrow 2\pi-\varphi, \; \alpha \rightarrow \alpha + \pi, \; \beta \rightarrow \pi - \beta, \; \gamma \rightarrow 2\pi - \gamma.
\end{equation}
Acceptable states must be eigenstates of the antisymmetrization operator $A = 1 - P_{23}$ with eigenvalue $2$, 
which implies the boundary conditions obeyed by the adiabatic eigenfunctions,  Eq.(\ref{eq:adiabaticeq}) giving the following spatial boundary conditions:
\footnotesize
\begin{align}
 J=0^{+}\,: \quad & \phi_{\nu} (R;\theta,0)=\phi_{\nu} (R;\theta,\pi)=0 \\
 J=1^{+}\,: \quad &\frac{\partial}{\partial\varphi}\phi_{\nu}(R;\theta,\varphi)\bigg\rvert_{\varphi=0}=\frac{\partial}{\partial\varphi}\phi_{\nu}(R;\theta,\varphi)\bigg\rvert_{\varphi=\pi}=0  \\
 J=1^{-}\,: \quad 
 &\frac{\partial}{\partial\varphi}\phi_{\nu c}(R;\theta,\varphi)\bigg\rvert_{\varphi=0}=\phi_{\nu c} (R;\theta,\pi)=0 \\
	&\phi_{\nu s} (R;\theta,0)=\frac{\partial}{\partial\varphi}\phi_{\nu s}(R;\theta,\varphi)\bigg\rvert_{\varphi=\pi}=0 \\
 J=2^{-}\,: \quad
 &\phi_{\nu c} (R;\theta,0)=\frac{\partial}{\partial\varphi}\phi_{\nu c}(R;\theta,\varphi)\bigg\rvert_{\varphi=\pi}=0 \\
	&\frac{\partial}{\partial\varphi}\phi_{\nu s}(R;\theta,\varphi)\bigg\rvert_{\varphi=0}=\phi_{\nu s} (R;\theta,\pi)=0 
\end{align}
\normalsize
For the system of three spin-up fermion, after applying the postsymmetrization operator $A=1-P_{12}-P_{23}-P_{31}+P_{123}+P_{132}$, the spatial boundary conditions can be represented as:
\footnotesize
\begin{align}
 J=0^{+}\,: \quad & \phi_{\nu} (R;\theta,0)=\phi_{\nu} (R;\theta,\pi/3)=0 \\
 J=1^{+}\,: \quad &\frac{\partial}{\partial\varphi}\phi_{\nu}(R;\theta,\varphi)\bigg\rvert_{\varphi=0}=\frac{\partial}{\partial\varphi}\phi_{\nu}(R;\theta,\varphi)\bigg\rvert_{\varphi=\pi/3}=0 \\
 J=1^{-}\,: \quad
 &\phi_{\nu c} (R;\theta,0)=\frac{\partial}{\partial\varphi}\phi_{\nu c}(R;\theta,\varphi)\bigg\rvert_{\varphi=\pi/3}=0 \\
	&\frac{\partial}{\partial\varphi}\phi_{\nu s}(R;\theta,\varphi)\bigg\rvert_{\varphi=0}=\phi_{\nu s} (R;\theta,\pi/3)=0 \\
 J=2^{-}\,: \quad 
 &\frac{\partial}{\partial\varphi}\phi_{\nu c}(R;\theta,\varphi)\bigg\rvert_{\varphi=0}=\phi_{\nu c} (R;\theta,\pi/3)=0 \\
	&\phi_{\nu s} (R;\theta,0)=\frac{\partial}{\partial\varphi}\phi_{\nu s}(R;\theta,\varphi)\bigg\rvert_{\varphi=\pi/3}=0 
\end{align}
\normalsize
Substitution of $\psi_{E}(R;\Omega)$ from Eq.(\ref{eq:totalwavefunc}) into the Schr\"{o}dinger equation Eq.(\ref{eq:Schrodingereq}) leads to a set of one dimensional coupled hyperradial differential equations
\begin{align}\label{eq:1dode}
	&\left[-\frac{1}{2\mu}\frac{d^{2}}{dR^{2}}+U_{\nu}(R)-E\right]F_{\nu E}(R) \nonumber\\
	&-\frac{1}{2\mu}\sum_{\nu'}\left[2P_{\nu\nu'}(R)\frac{d}{dR}+Q_{\nu\nu'}(R)\right]F_{\nu' E}(R)=0
\end{align}
In the above expression, $E$ is the total energy and  $W_{\nu}(R)$ is the effective {\it adiabatic potential} in channel $\nu$:
\begin{equation}
	W_{\nu}(R) \equiv U_{\nu}(R)-\frac{1}{2\mu}Q_{\nu\nu}(R)
\end{equation}
The non-adiabatic coupling matrices $P_{\nu\nu'}(R)$ and $Q_{\nu\nu'}(R)$ are defined as 
\begin{align}
	&P_{\nu\nu'}(R)=\int d\Omega \Phi_{\nu}^{*}(R;\Omega)\frac{\partial}{\partial R}\Phi_{\nu'}(R;\Omega) \\
	&Q_{\nu\nu'}(R)=\int d\Omega \Phi_{\nu}^{*}(R;\Omega)\frac{\partial^{2}}{\partial R^{2}}\Phi_{\nu'}(R;\Omega)
\end{align} 
The radial Eqs.(\ref{eq:1dode}) can be solved by $R$-matrix propagation.\cite{wang2011PRA} The effective adiabatic potentials are represented asymptotically by their behavior at $(R\rightarrow \infty)$, namely:
\begin{equation}\label{eq:wasym}
	W_{\nu}(R)=\frac{l_{e}(l_{e}+1)}{2\mu R^{2}}
\end{equation}
where $l_{e}$ controls the effective angular momentum barrier of the three free asymptotic particles in the large hyperradius limit, $R\rightarrow \infty$.
%%%%%%%%%%%%%%%%%%%%%%%%%%%%%%%%%%%%%%%%%%%%%%%%%%%%
\section{Results}

%%%%%%%%%%%%%%%%%%%%%%%%%%%%%%%%%%%%%%%%%%%%%%%%%%%%
\begin{table}[h]
\caption{Comparison of our present results with those obtained by Werner and Castin\cite{werner2006unitary} and by Blume {\it et al.}\cite{blume2007universal} for two symmetries $J^{\Pi}=0^{+}$ and $J^{\Pi}=1^{-}$ in the 
three-fermion system ($\downarrow\uparrow\uparrow$). $l_{e}$ is the effective angular momentum and $l_{e}(l_{e}+1)$ is the coefficient of $1/(2\mu R^{2})$ (See Eq.(\ref{eq:wasym})). The script "$1^{\text{st}}$ $s$-wave", "$2^{\text{nd}}$ $s$-wave" and "$3^{\text{rd}}$ $s$-wave" are represented as the first $s$-wave unitary limit, the second $s$-wave unitary limit and the third $s$-wave unitary channel, respectively. The value of $E_{\nu n}$ represents the total relative energy, which is related to the $l_e$ value through $E_{\nu n}=(l_e^{(\nu)} + 2n +3)\hbar\omega$. $E_{00}$ is the (relative) ground state of the Fermi system of 3 atoms in an isotropic harmonic trap, $E_{10}$ and $E_{20}$ are the first and second excited energy levels, respectively.}
\footnotesize
\begin{tabular}{c|ccc|ccc}
\hline\hline
     &.    & $J^{\Pi}=0^{+}$&           &     & $J^{\Pi}=1^{-}$& \\
\hline 
  & Present & Ref.\cite{blume2007universal} & Ref.\cite{werner2006unitary} & Present & Ref.\cite{blume2007universal} & Ref.\cite{werner2006unitary} \\
\hline
 \footnotesize	 $1^\text{st}$ $s$-wave ($l_e^{(\nu)}$) & 1.666  & 1.682 & 1.666 & 1.272 & 1.275 & 1.272   \\
 $E_{00}$/($\hbar \omega$)  & 4.666  & 4.682 & 4.666 & 4.272 & 4.275 & 4.272        \\
\hline
 \footnotesize	 $2^\text{nd}$ $s$-wave ($l_e^{(\nu)}$) & 4.628  & 4.637 & 4.627 & 3.861 & 3.868 & 3.858   \\
 $E_{10}$/($\hbar \omega$)  & 7.628  & 7.637 & 7.627 & 6.861 & 6.868 & 6.858        \\
\hline
 \footnotesize	 $3^\text{rd}$ $s$-wave ($l_e^{(\nu)}$) & 6.615  & 6.628 & 6.614 & 5.215 & 5.229 & 5.216   \\
 $E_{20}$/($\hbar \omega$)  & 9.615  & 9.628 & 9.614 & 8.215 & 8.229 & 8.216        \\
\hline\hline
\end{tabular}
\label{table:comparele}
\end{table}
\normalsize
%%%%%%%%%%%%%%%%%%%%%%%%%%%%%%%%%%%%%%%%%%%%%%%%%%%%

\subsection{$s$-wave universal properties in the three-body potential}
\begin{figure}[h]
\includegraphics[width=9cm]{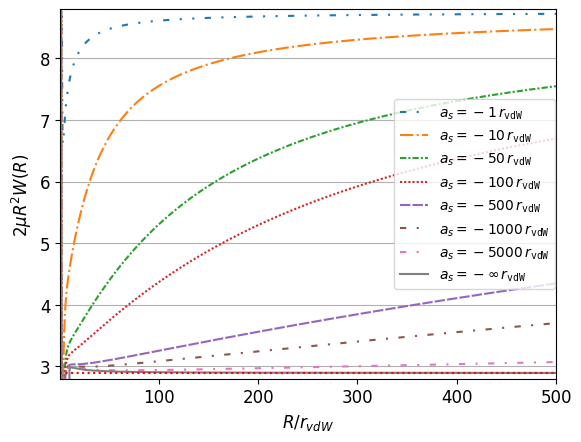}
\caption{(color online). Shown are the lowest adiabatic potential curves versus hyperradius for several different $s$-wave scattering lengths ($a_{s}$) in the van der Waals length unit $r_{\text{vdW}}$, for the system ($\downarrow\uparrow\uparrow$) with $J^{\Pi}=1^{-}$. As $|a_{s}|$ gets larger, the potential curve should approach the red dotted line whose value equals the coefficient of $1/(2\mu R^{2})$ in the adiabatic potential energy curve.}
\label{fig:potandas}
\end{figure}

The adiabatic equation Eq.(\ref{eq:adiabaticeq}) is solved for the effective adiabatic potential energy curves for various scattering lengths. For example, the Fig.\ref{fig:potandas} presents the solutions of the Eq.(\ref{eq:adiabaticeq}), and demonstrates that $2\mu R^{2} W_{\nu}(R)$ is a constant throughout the range from $R=100\ r_{\text{vdW}}$ to at least $R=500\ r_{\text{vdW}}$. The quantity $r_{\text{vdW}}$ is the van der Waals length and is defined as $r_{\text{vdW}}\equiv\frac{1}{2}\left(2\mu C_{6}/\hbar^{2}\right)^{1/4}$. The energy scale in our calculations is the van der Waals energy unit, defined as $E_{\text{vdW}}=\hbar^{2}/(2\mu_{2b}r_{\text{vdW}}^{2})$. In this regime, the adiabatic potential has a stable asymptotic coefficient of $1/(2\mu R^{2})$, represented here either as $l_e(l_e+1)$ for repulsive potentials or else as $s_0^2-1/4$ with an imaginary value of $s_0$ for attractive potentials. Our results are also compared to the results of Werner and Castin\cite{werner2006unitary} and of Blume {\it et al.}\cite{blume2007universal}, who treated the few-fermion problem in an isotropic harmonic trap at the unitary limit.  As Table.\ref{table:comparele} demonstrates, our results agree well with those previous results, even though we used a different two-body model interaction, namely the Lennard-Jones potential;  adjustment of the parameter $\lambda_n$ allows us to take the limit of infinite $s$-wave scattering length.  This confirms the universal nature of these properties.  Our study next explores a different situation, where the interaction between the two spin-up particles is taken to the $p$-wave unitary limit for different symmetries $J^{\Pi}$. 

%%%%%%%%%%%%%%%%%%%%%%%%%%%%%%%%%%%%%%%%%%%%%%%%%%%%
%%%%%%%%%%%%%%%%%%%%%%%%%%%%%%%%%%%%%%%%%%%%%%%%%%%%%

\subsection{$p$-wave universality for a system with two spin-up fermions and one spin-down fermion} \label{sec:spunitary}
\subsubsection{Universal $p$-wave interaction between same spin fermions only}\label{sec:sspunitary}
In Fig.\ref{fig:sps}, the interaction between spin-up and spin-down particles is fixed at the $s$-wave unitarity (which also implies a $p$-wave scattering volume equal to $V_p=-1.974 \, r_{\text{vdW}}^3$) limit, and the interaction between fermions in the same spin state is fixed at the $p$-wave unitary limit with total symmetry $J^{\Pi}=0^{+},\, 1^{+}, \, 1^{-}$ and $2^{-}$.
One possibly surprising result is our finding that many of the asymptotic $s$-wave unitary potentials remain unchanged when the $p$-wave interaction is tuned to infinity, but with the emergence of new additional potentials at the $p$-wave unitarity, having values of $l_e$ that did not exist for the pure $s$-wave case.
In Fig.(\ref{fig:spsJ0p}) and Fig.(\ref{fig:spsJ1m}), the novel additional potential curves associated with the $p$-wave unitary channels (solid curve for $J^{\Pi}=0^{+}$, and solid and dashdot curves for $J^{\Pi}=1^{-}$) are obtained for in this situation. The term {\it $p$-wave unitary channel} means that when the $p$-wave scattering volume between one or more pairs of particles is increased to infinity, this is an entirely new channel that emerges, having a value of $l_e$ different from the usual non-interacting values of $l_e$, which is the hallmark of Efimov-related physics.  Despite the additional potentials just described, we still see the emergence of $s$-wave unitary channels where $l_e(l_e+1)$ (i.e., the coefficients of $1/(2\mu R^{2})$ of the asymptotic $s$-wave channels) are close to the results of Werner and Castin and of Blume {\it et al.}. Similarly, {\it $s$-wave unitary channels} are defined to be those that have $l_e$ modified from their non-interacting values, when the $s$-wave scattering length between one or more pairs of particles increases to infinity. The terminology {\it NI asymptotically} means that a channel has the same $l_e$ value as the non-interacting channel at asymptotic hyperradii. According to our numerical results, the adiabatic 3-body potentials of the additional $p$-wave unitary channels asymptotically approach constant coefficients of $1/(2\mu R^{2})$. For the lowest such $p$-wave unitary channels, the coefficient $l_e(l_e+1)$ is close to the value $2.00$ for the symmetry $J^{\Pi}=0^{+}$ and approaches the values $0.00$ and $6.00$ for the two $p$-wave unitary channels in the symmetry $J^{\Pi}=1^{-}$. From our calculations, additional $p$-wave unitary channels are predicted to occur in the first channel for $J^{\Pi}=0^{+}$, and in the first and third channels for $J^{\Pi}=1^{-}$. Unlike the $s$-wave unitary channels, there are apparently only two $p$-wave unitary channels, rather than an infinity of such channels with modified centrifugal potentials. In Fig.(\ref{fig:spsJ1p}) and Fig.(\ref{fig:spsJ2m}), angular momentum selection rules prevent the existence of an $s$-wave for either Jacobi vector, and therefore there are no  $s$-wave unitary channels (i.e. with modified $l_e$)  for those parity-unfavored symmetries. However, we still find one $p$-wave unitary channel in each of these two parity-unfavored cases $1^+$ and $2^-$.  The coefficients of $1/(2\mu R^{2})$ are extremely stable asymptotically, approaching values equal to $2.00$ and $6.00$, for total angular momenta and parities $J^{\Pi}=1^{+}$ and $J^{\Pi}=2^{-}$, respectively. 

\twocolumngrid

\subsubsection{Universal $p$-wave interaction between opposite spin fermions only}\label{sec:pspunitary}
Fig.\ref{fig:psp} shows potential curves relevant for four different symmetries in three-body systems relevant to the two-component Fermi gas, at the  $p$-wave unitary limit ($V_{p}\rightarrow \infty$ and $a_s = 1.987 r_{\text{vdW}}$, at the first $p$-wave pole of the Lennard-Jones potential) between two fermions in different spin states. For each of these four symmetries, the $p$-wave interaction between the fermions in identical spin states has been chosen to be comparatively weak, namely set to $V_{p}=-2\ r_{\text{vdW}}^3$.  %with different total angular momentum. 
In Fig.(\ref{fig:pspJ0p}) and Fig.(\ref{fig:pspJ1p}), a single $p$-wave unitary channel has been found for these symmetries $0^+$ and $1^+$, namely a coefficient of $1/(2\mu R^{2})$ approaching $2.01$ asymptotically for each of these two distinct symmetry cases, $J^\Pi=0^+$ and $1^+$, respectively.  Fig.(\ref{fig:pspJ1m}) considers the symmetry $1^-$, where two $p$-wave unitary potential curves (or channels) are observed.  The coefficients of $1/(2\mu R^{2})$ are very close to the results obtained for the case considered in subsection (\ref{sec:sspunitary}), where the $p$-wave interaction between like spin fermions was set to unitarity for the symmetry $J^{\Pi}=1^{-}$ (See Fig.(\ref{fig:spsJ1m})). The values of the asymptotic centrifugal coefficients are approximately $\approx 0.00$ and $\approx 6.01$ for the first and second adiabatic potential curves, respectively. These values are the same (to within our numerical accuracy) as the values of $l_e(l_e+1)$ for these two symmetries in \ref{sec:sspunitary}.  In Fig.(\ref{fig:pspJ2m}), the $p$-wave unitary channel achieves a stable asymptotic behavior, and the coefficient of $1/(2\mu R^{2})$ is found to be close to $6.00$ asymptotically. 

\subsubsection{Universal $p$-wave interaction between same spin fermions and opposite spin fermions}
\begin{figure}[h]
\subfloat[$J^{\Pi}=0^{+}$\label{fig:2FpppJ0p}]
{\includegraphics[width=8cm]{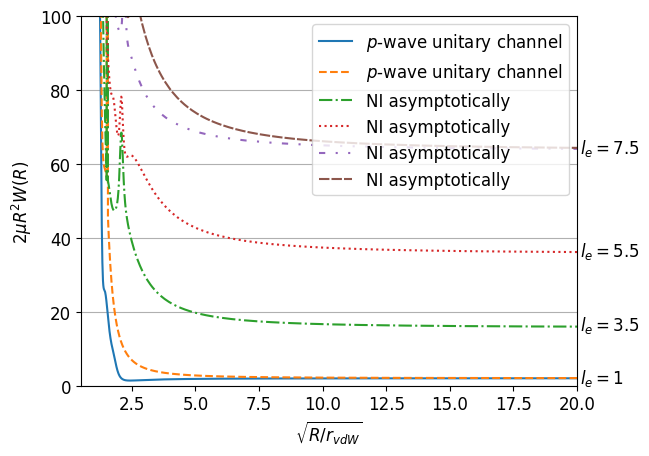}}\centering
\subfloat[$J^{\Pi}=1^{+}$\label{fig:2FpppJ1p}]
{\includegraphics[width=8cm]{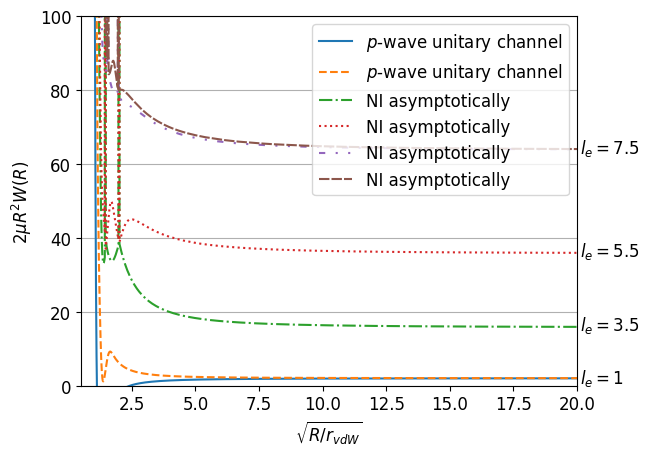}} \\
\subfloat[$J^{\Pi}=1^{-}$\label{fig:2FpppJ1m}]
{\includegraphics[width=8cm]{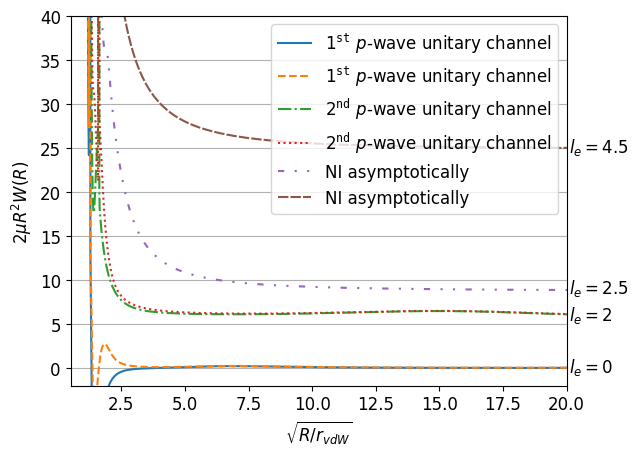}}\centering
\subfloat[$J^{\Pi}=2^{-}$\label{fig:2FpppJ2m}]
{\includegraphics[width=8cm]{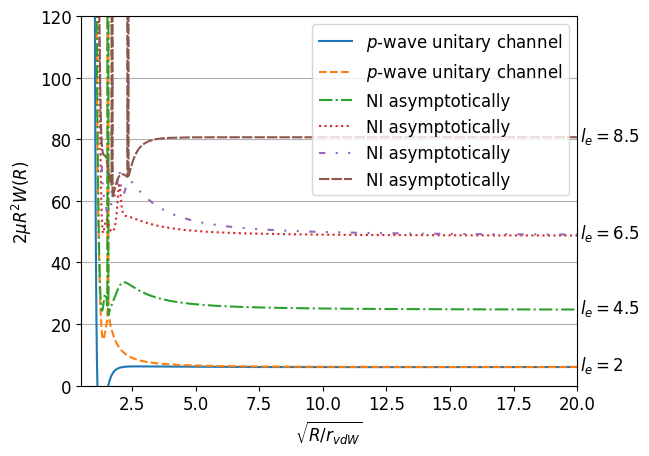}}
\caption{(color online).  Shown are the adiabatic potential curves for the first to sixth channels versus hyperradius for the fermions in the same and different spin state both at the $p$-wave unitary limit ($V_{p} \rightarrow \infty$) for the various symmetries $J^{\Pi}$. (a) and (b), the solid (blue) and dashed (orange) curve corresponds to the $p$-wave degenerate unitary channel and the coefficient of $1/(2\mu R^{2})$ is approximately $2$. (c), the solid (blue) curve and dashed (orange) curves show the $1^{\text{st}}$ degenerate $p$-wave universal property and their coefficients of $1/(2\mu R^{2})$ are computed here to have the values $\approx 0.01$. The dash-dotted (green) curve and dotted (red) curve represent the $2^{\text{nd}}$ degenerate $p$-wave unitary channels which coefficients of $1/(2\mu R^{2})$ are closed to $6$. (d), the two lowest adiabatic potential curves (solid and dashed curves) represent the degenerate $p$-wave unitary channels and the $1/(2\mu R^{2})$ coefficient is calculated to have the value $\approx 6$.}
\label{fig:2Fppp}
\end{figure}

In Fig.\ref{fig:2Fppp}, the fermion interactions in both the same and different spin states are set at the $p$-wave unitary limit ($V_p\rightarrow \infty$ and $a_s = 1.987\,r_{\text{vdW}}$, for the first $p$-wave pole of the Lennard-Jones potential) with multiple symmetries $J^\Pi = 0^+, \, 1^+, \, 1^-$ and $2^-$. In this situation, there arises a combination of the above two cases. The $p$-wave unitary channels exhibit degeneracies in these four symmetries. In Fig.(\ref{fig:2FpppJ0p}) and Fig.(\ref{fig:2FpppJ1p}),  two $p$-wave unitary channels emerge with the symmetry $J^\Pi=0^+$ and $1^+$, with their asymptotic coefficient of $1/(2\mu R^2)$ close to $2.00$, implying that the $l_e$ values are close to $1.00$. In Fig.(\ref{fig:2FpppJ1m}), the $p$-wave unitary channels are doubly-degenerate and the $1/(2\mu R^2)$ coefficients are found to be approximately $0.01$ and $6.06$. In Fig.(\ref{fig:2FpppJ2m}), there are degenerate $p$-wave unitary channels whose coefficients of $1/(2\mu R^2)$ are found numerically to equal $6.01$.

\twocolumngrid

\subsection{$p$-wave universality with three spin-polarized fermions}
\begin{figure}[h]
\subfloat[$J^{\Pi}=0^{+}$\label{fig:pppJ0p}]
{\includegraphics[width=8cm]{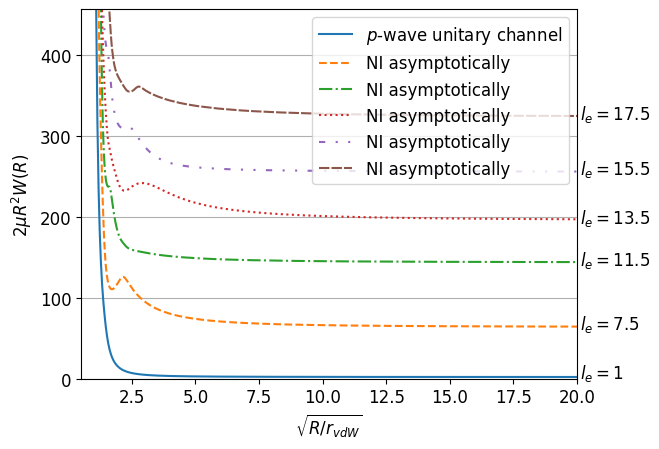}}\centering
\subfloat[$J^{\Pi}=1^{+}$\label{fig:pppJ1p}]
{\includegraphics[width=8cm]{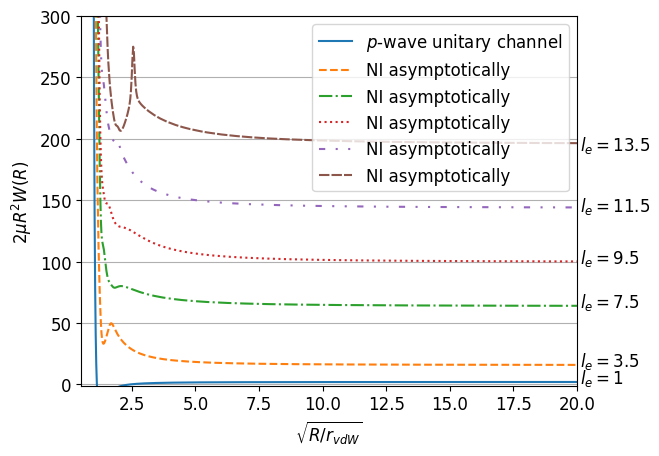}} \\
\subfloat[$J^{\Pi}=1^{-}$\label{fig:pppJ1m}]
{\includegraphics[width=8cm]{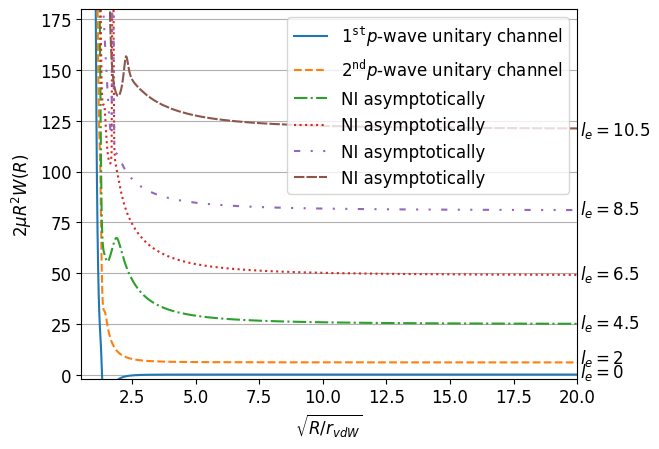}}\centering
\subfloat[$J^{\Pi}=2^{-}$\label{fig:pppJ2m}]
{\includegraphics[width=8cm]{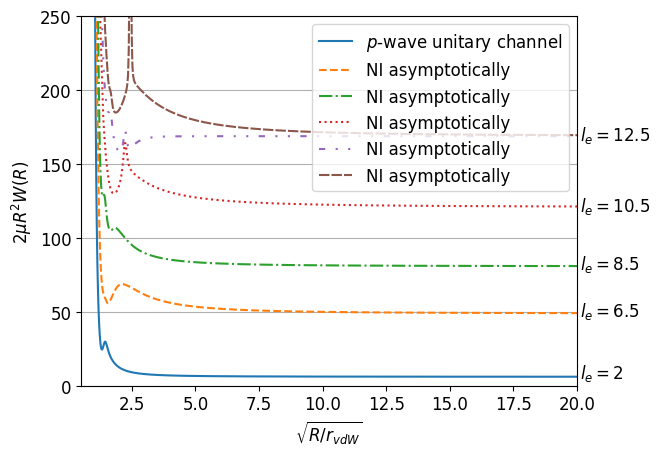}}
\caption{(color online). Shown are the adiabatic potential curves for the first to sixth channels versus hyperradius for the single-component 3-fermion system ($\uparrow \uparrow \uparrow $) with their two-body interactions set at the $p$-wave unitary limit ($V_{p} \rightarrow \infty$) for the various symmetries $J^{\Pi}$. (a) and (b), the solid (blue) curve corresponds $p$-wave unitary channel and the coefficient of $1/(2\mu R^{2})$ is approximately $2$. (c), the solid (blue) and the dashed (orange) curve show the $p$-wave universal property, with asymptotic coefficients of $1/(2\mu R^{2})$ computed here to have the values $\approx 0.01$ and $\approx 6$, respectively. (d), the lowest adiabatic potential curve (solid curve) represents the $p$-wave unitary channel and the $1/(2\mu R^{2})$ coefficient is calculated to have the value $\approx 6$.}
\label{fig:ppp}
\end{figure}

Fig.\ref{fig:ppp} shows our 3-body calculation for a single-component Fermi gas at the $p$-wave unitarity limit. In Fig.(\ref{fig:pppJ0p}) and Fig.(\ref{fig:pppJ1p}), the coefficient of $1/(2\mu R^{2})$ of the lowest adiabatic potential curve (blue curve) is close to the integer $2$ in the two symmetries $J^{\Pi}=0^{+}$ and $1^{+}$. In Fig.(\ref{fig:pppJ1m}), one can see two stable $p$-wave universal adiabatic potential curves (blue and orange curves) with total orbital angular momentum $J^{\Pi}=1^{-}$, whose coefficients of $1/(2\mu R^{2})$ in the novel $p$-wave unitary channels are close to the integer values $0$ and $6$. In Fig.(\ref{fig:pppJ2m}), the $p$-wave unitary channel shows a stable asymptotic coefficient of $1/(2\mu R^{2})$ equal to $6$, for the symmetry $J^{\Pi}=2^{-}$.\\

According to the above results, the $p$-wave unitary channel is ubiquitous in the three-body adiabatic potential curves and apparently its existence would not vary for different spin state cases.  This means that 3 fermions in two different spin states and three identical spin fermions both have $p$-wave unitary channels and behavior.  The $p$-wave unitary channels are shown to have very stable asymptotic effective centrifugal potentials, in all cases with a repulsive (or vanishing) coefficient. The coefficients of $R^{-2}$ are very close for the three body systems with all fermions in the same spin state ($\uparrow\uparrow\uparrow$), and for the case of two identical fermions and a distinguishable particle  ($\uparrow\uparrow\downarrow$) with total orbital angular momenta $J^{\Pi}=0^{+}, 1^{+}, 1^{-}$ and $2^{-}$ at the $p$-wave unitary limit. 

\newpage

\onecolumngrid

%%%%%%%%%%%%%%%%%%%%%%%%%%%%%%%%%%%%%%%%%%%%%%%%%%%%%
\begin{figure}[h]
\subfloat[$J^{\Pi}=0^{+}$\label{fig:spsJ0p}]
{\includegraphics[width=8cm]{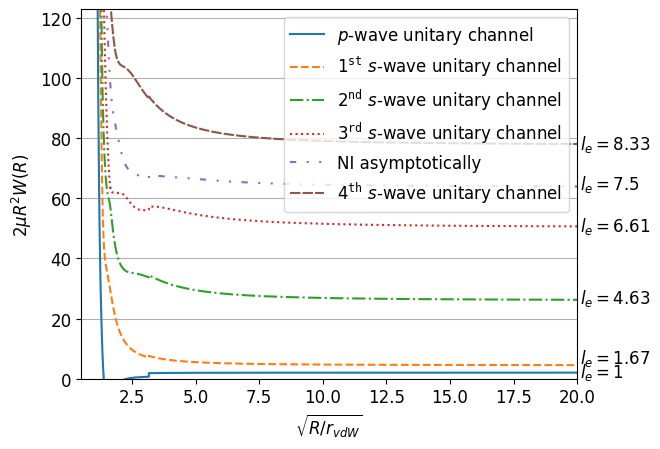}}\centering
\subfloat[$J^{\Pi}=1^{+}$\label{fig:spsJ1p}]
{\includegraphics[width=8cm]{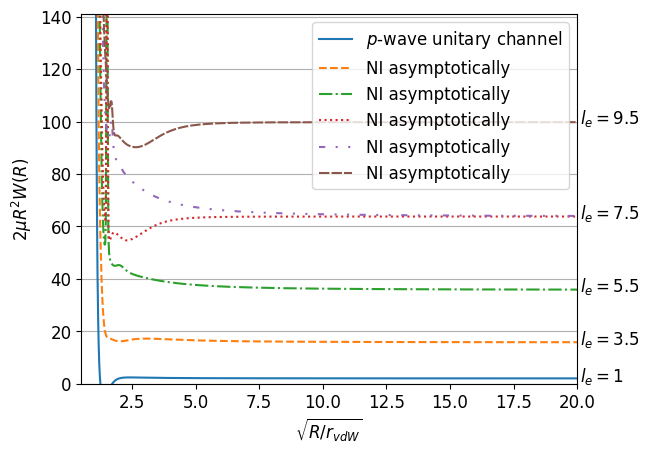}}\\
\subfloat[$J^{\Pi}=1^{-}$\label{fig:spsJ1m}]
{\includegraphics[width=8cm]{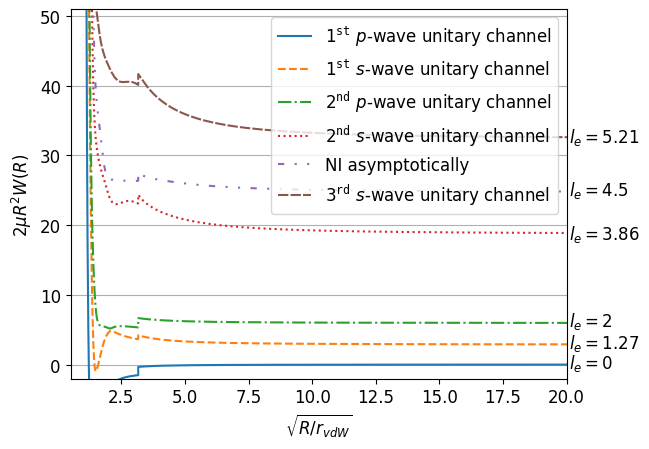}}\centering
\subfloat[$J^{\Pi}=2^{-}$\label{fig:spsJ2m}]
{\includegraphics[width=8cm]{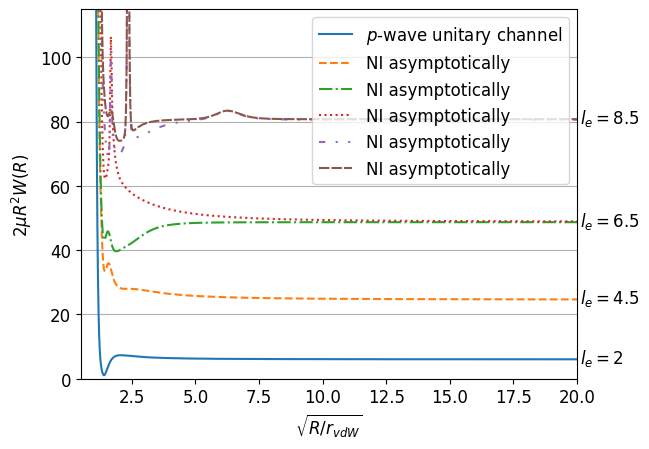}}
\caption{(color online). Shown are the adiabatic potential curves for the first to sixth channels versus hyperradius for the ($\downarrow\uparrow\uparrow$) system.  The interaction between the two fermions in the  same spin state has been set at the $p$-wave unitary limit ($V_{p} \rightarrow \infty$), and the interactions between fermions in different spin states have also been set at the $s$-wave unitary limit ($a_{s} \rightarrow \infty$), for the various symmetries $J^{\Pi}$. (a) and (b), the solid (blue) curve corresponds to the $p$-wave unitary channel and the coefficient of $1/(2\mu R^{2})$ is approximately $2$. (c), the solid (blue) curve and dash-dot (green) curve shows $p$-wave universal property and their coefficients of $1/(2\mu R^{2})$ are computed here to have the values $\approx 0.02$ and $\approx 6$, respectively. (d), the lowest adiabatic potential curve (solid curve) represents the $p$-wave unitary channel and the asymptotic $1/(2\mu R^{2})$ coefficient is calculated to have the value $\approx 6$.}
\label{fig:sps}
\end{figure}
%%%%%%%%%%%%%%%%%%%%%%%%%%%%%%%%%%%%%%%%%%%%%%%%%%%%%

\onecolumngrid

%%%%%%%%%%%%%%%%%%%%%%%%%%%%%%%%%%%%%%%%%%%%%%%%%%%%%
\begin{figure}[h]
\subfloat[$J^{\Pi}=0^{+}$\label{fig:pspJ0p}]
{\includegraphics[width=8cm]{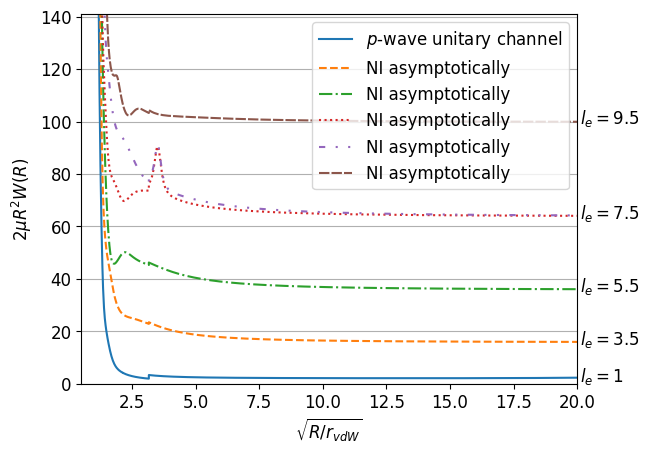}}\centering
\subfloat[$J^{\Pi}=1^{+}$\label{fig:pspJ1p}]
{\includegraphics[width=8cm]{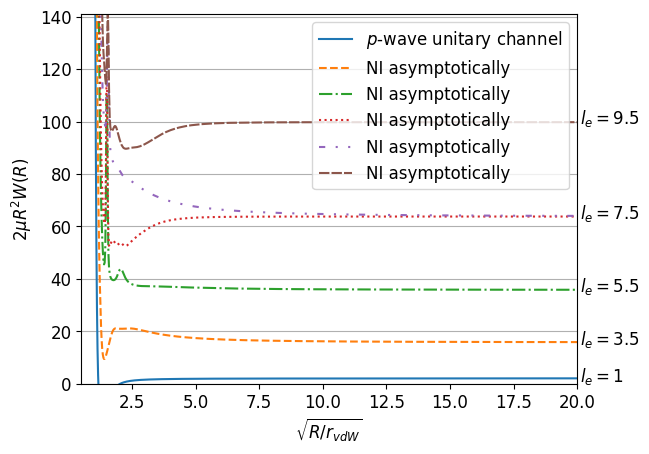}} \\
\subfloat[$J^{\Pi}=1^{-}$\label{fig:pspJ1m}]
{\includegraphics[width=8cm]{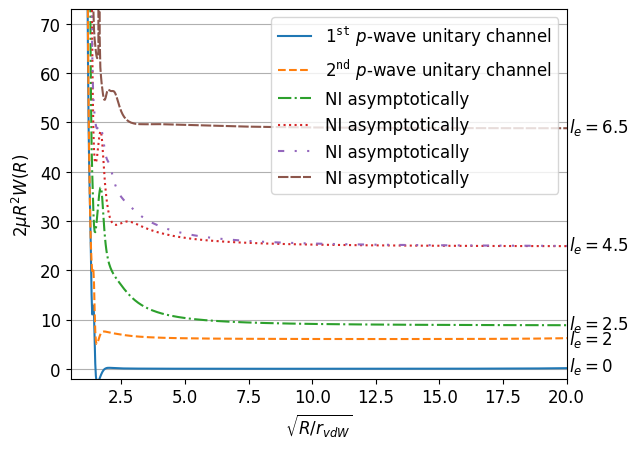}}\centering
\subfloat[$J^{\Pi}=2^{-}$\label{fig:pspJ2m}]
{\includegraphics[width=8cm]{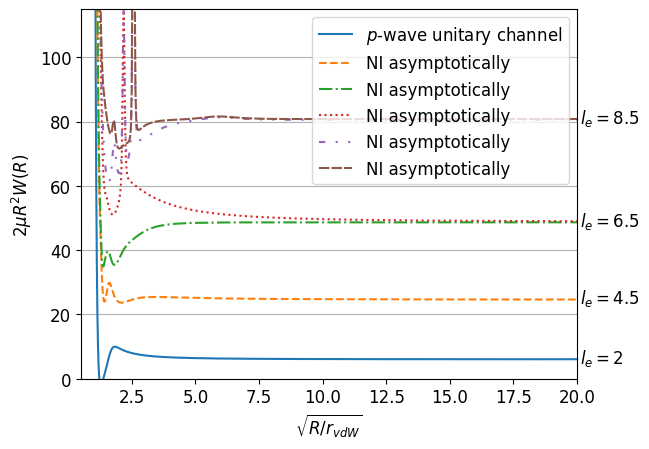}}
\caption{(color online). Shown are the adiabatic potential curves for the first to sixth channels versus hyperradius for the two-component 3-fermion system ($\downarrow\uparrow\uparrow$), with the interaction between fermions in different spin states set at the $p$-wave unitary limit ($V_{p} \rightarrow \infty$) and the interaction between the same spin state fermions set at a weak value with $p$-wave scattering volume close to $V_{p}=-2\, r_{\text{vdW}}^{3}$, for several symmetries $J^{\Pi}$. (a) and (b), the solid (blue) curve corresponds to the $p$-wave unitary channel and the coefficient of $1/(2\mu R^{2})$ is approximately $2$. (c), the solid (blue) curve and dashed (orange) curve exhibit the $p$-wave universal property, and their asymptotic coefficients of $1/(2\mu R^{2})$ are computed here to have the values $\approx 0$ and $\approx 6$, respectively. (d), the lowest adiabatic potential curve (solid curve) represents the $p$-wave unitary channel and the asymptotic $1/(2\mu R^{2})$ coefficient is calculated to have the value $\approx 6$.}
\label{fig:psp}
\end{figure}
%%%%%%%%%%%%%%%%%%%%%%%%%%%%%%%%%%%%%%%%%%%%%%%%%%%%%

\onecolumngrid

%%%%%%%%%%%%%%%%%%%%%%%%%%%%%%%%%%%%%%%%%%%%%%%%%%%%%
%%%%%%%%%%%%%%%%%%%%%%%%%%%%%%%%%%%%%%%%%%%%%%%%%%%%

\onecolumngrid

%%%%%%%%%%%%%%%%%%%%%%%%%%%%%%%%%%%%%%%%%%%%%%%%%%%%%
%%%%%%%%%%%%%%%%%%%%%%%%%%%%%%%%%%%%%%%%%%%%%%%%%%%%%

\twocolumngrid

\subsection{On the nonexistence of a $p$-wave Efimov effect}
In Fig.(\ref{fig:pspJ0p}), Fig.(\ref{fig:pspJ1p}) and Fig.(\ref{fig:pspJ1m}), our results demonstrate that the $p$-wave Efimov effect cannot be observed in these symmetries. These results disagree with the results suggested by Braaten {\it et al.}\cite{braaten2012renormalization} It should be pointed out that our study uses a different two-body interaction potential, and that these results can be affected by the $p$-wave effective range in any particular system.  The two-body system has a nonzero but short range potential, whereby the $p$-wave phase shift, scattering volume, and effective range can be written at $k\rightarrow 0$ as~\cite{pricoupenko2006pseudopotential}
\begin{equation}
	k^{3}\cot\delta_{1}(k) \rightarrow -\frac{1}{V_{p}}+\frac{1}{2}r_{p}k^{2}+O(k^{4})
\end{equation}
where $\delta_{1}$ is the $p$-wave phase shift, $V_{p}$ is the $p$-wave scattering volume and $r_{p}$ is the $p$-wave ``effective range''. The units of the $p$-wave scattering volume and $p$-wave effective range are $(\text{length})^{3}$ and $1/(\text{length})$, respectively. When the $p$-wave scattering volume approaches infinity, the $p$-wave effective range would be closed to $r_{p}=-1.7\, r_{\text{vdW}}^{-1}$ in the van der Waals tail~\cite{gao1998quantum}. Surprisingly, if we would utilize the Born-Oppenheimer approximation to our adiabatic potential curves, i.e. incorrectly neglecting the diagonal elements of the $Q$ matrix, such a treatment would erroneously seem to imply the existence of a $p$-wave Efimov effect in the lowest adiabatic potential curve with $J^{\Pi}=1^{-}$ symmetry. 

The nonexistence of a true Efimov effect, which occurs whenever the centrifugal coefficient is more negative than $-1/4$, is by now well documented, but in different theoretical treatments the reason for that nonexistence can look quite different. In effective field theory treatments, for instance, it was proven that a three-body wavefunction that {\it initially appears} to support an Efimov effect is in fact incorrect and would have negative probabilities~\cite{nishida2012impossibility}.  In the hyperspherical picture, on the other hand, it is crucial to look at the asymptotic potential curves that include the diagonal $Q$ elements.  We have observed a number of cases where the potential curves neglecting $Q$ would appear to indicate the presence of an Efimov effect, but in all cases checked so far involving $p$-wave interactions at unitarity and equal mass particles, once $Q$ is included there is no Efimov effect. 
Specifically our calculations show no evidence of any $p$-wave Efimov effect when $V_{p}$ goes to infinity, for any combination of spin-up and spin-down fermions with the following symmetries: $J^{\Pi}=0^{+}, 1^{+}$, $1^{-}$ and $2^{-}$. Similar results are found for a three-body system having two spin-up fermions interacting at the $p$-wave unitary limit and the opposite spin fermions near $s$-wave unitarity in the symmetry $J^{\Pi}=1^{-}$ (See Fig.(\ref{fig:spsJ1m})). The Born-Oppenheimer potential (neglecting $Q$) shows (incorrectly) a $p$-wave Efimov effect since the lowest three-body potential curve would then be asymptotically attractive with the coefficient of $1/(2\mu R^{2})$ close to $-1.06$. However, the full adiabatic potential curve that includes the added diagonal $Q_{\nu\nu}(R)$ matrix element does not exhibit any evidence of a $p$-wave Efimov effect, and it is the adiabatic potential curve that matters. 

Moreover, the same issue discussed above arises for three spin-polarized fermions.  Fig.(\ref{fig:pppJ1m}) demonstrates that the $p$-wave Efimov effect would {\it incorrectly} seem to arise within the strict Born-Oppenheimer approximation that neglects $Q$.  However, our results show consistently that no $p$-wave Efimov effect occurs in the true adiabatic potential curves that include the diagonal elements of $Q$.  

\onecolumngrid

%%%%%%%%%%%%%%%%%%%%%%%%%%%%%%%%%%%%%%%%%%%%%%%%%%%%%
\begin{figure}[h]
\subfloat[$\uparrow\downarrow$ at $p$-wave unitary limit\label{fig:pspJ1mUW}]
{\includegraphics[width=8cm]{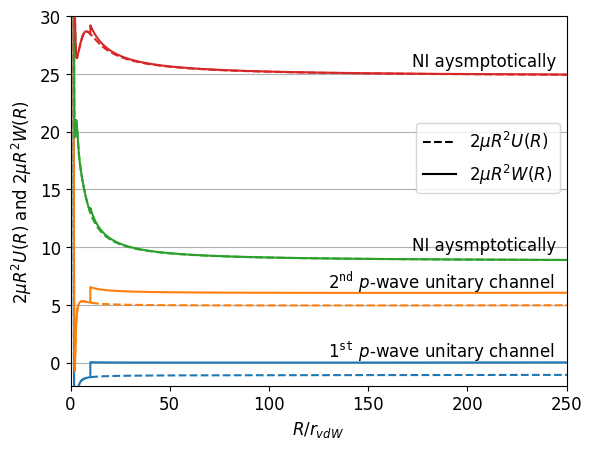}}\centering
\subfloat[$\uparrow\downarrow$ at $s$-wave unitary and $\uparrow\uparrow$ at $p$-wave unitary limit\label{fig:spsJ1mUW}]
{\includegraphics[width=8cm]{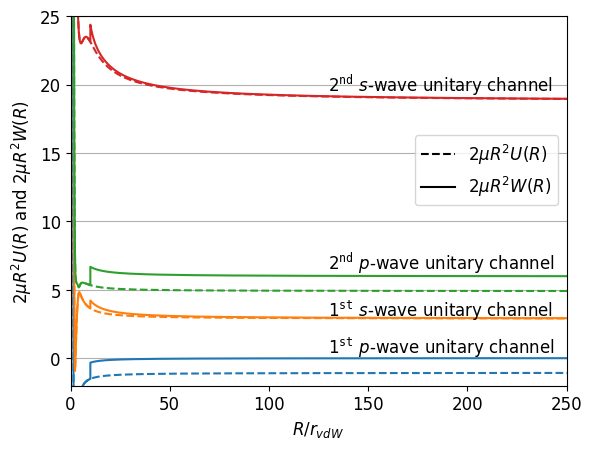}}\\
\subfloat[$\downarrow\uparrow\uparrow$ all at $p$-wave unitary limit\label{fig:2FpppJ1mUW}]
{\includegraphics[width=8cm]{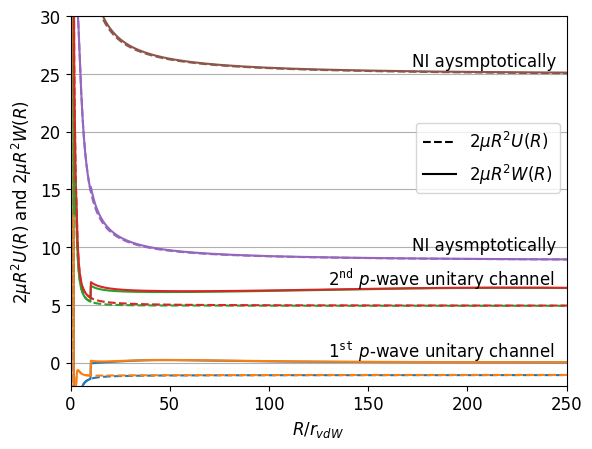}}\centering
\subfloat[$\uparrow\uparrow\uparrow$ at $p$-wave unitary limit\label{fig:pppJ1mUW}]
{\includegraphics[width=8cm]{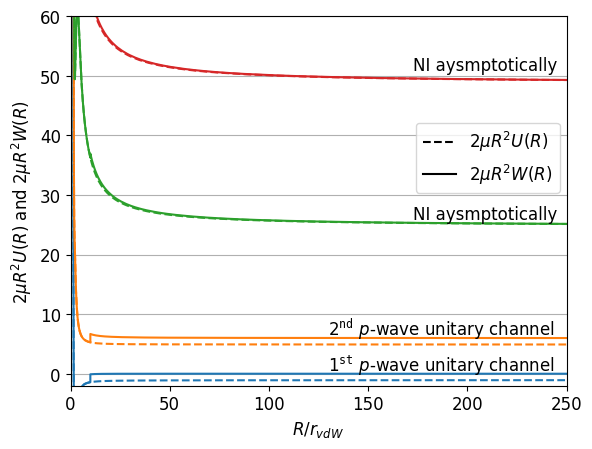}}
\caption{(color online). Comparison of the first to fourth channel potentials in the Born-Oppenheimer approximation (dashed curves) and  in the adiabatic approximation (solid curves) versus hyperradius in the symmetry $J^{\Pi}=1^{-}$. The $Q_{\nu\nu}(R)$ diagonal element has been  added to the Born-Oppenheimer potential only beyond $R=10\, r_{\text{vdW}}$ in order to see its strength;  it does not vanish asymptotically in the $p$-wave unitary channels faster than $1/R^2$ and is in fact asymptotically proportional to $1/R^{2}$. 
In (a), the spin-up and spin-down fermions close to the $p$-wave unitary limit and fermions in the same spin state have a weak interaction that produces a scattering volume equal to $V_{p}=-2\, r_{\text{vdW}}^{3}$. In (b), the fermions in the different spin states are close to the $s$-wave unitarity limit and the fermions in the same spin state have an interaction set at the $p$-wave unitary limit. (c), each pair of the fermions, i.e. both in the same or both in different spin states have an interaction at the $p$-wave unitary limit. (d), each pair out of the three spin-polarized fermions have an interaction at the $p$-wave unitarity limit.}
\label{fig:spsandpspUW}
\end{figure}
%%%%%%%%%%%%%%%%%%%%%%%%%%%%%%%%%%%%%%%%%%%%%%%%%%%%%
\vskip 0.3in

\twocolumngrid

\subsection{Crucial role of the diagonal $Q_{\nu\nu}(R)$ adiabatic correction, and universal behavior at the $p$-wave unitary limit}
The results shown in Fig.(\ref{fig:spsandpspUW}) demonstrate that the diagonal adiabatic correction elements $Q_{\nu\nu}(R)$  play a key role in determining the physically relevant effective adiabatic potential, asymptotically as well as at finite hyperradii. According to our calculations, the $Q_{\nu\nu}(R)$ matrix elements are proportional to $1/R^{2}$ in the $p$-wave unitary channels but are proportional to the $1/R^{3}$ in the $s$-wave unitary channels or the NI asymptotic channels. Thus the Born-Oppenheimer potentials without the diagonal correction included would seem to imply the existence of an Efimov effect with a negative coefficient of $1/R^{2}$ asymptotically.  The $Q_{\nu\nu}(R)$ qualitatively changes this in the more physically relevant adiabatic potentials, since in some cases they can change the coefficients of $1/(2\mu R^{2})$. The $p$-wave Efimov effect is in fact destroyed once the $Q_{\nu\nu}(R)$ matrix correction is added onto the lowest Born-Oppenheimer potential, because it causes the lowest adiabatic potential curve to either be repulsive asymptotically or else to have a near-zero coefficient of $1/R^{2}$.  

Fig.(\ref{fig:pspJ1mUW}) considers the case where different-component fermions interact at the $p$-wave unitary limit whereas the fermions in the same spin state experience a weak $p$-wave interaction, for the symmetry $J^{\Pi}=1^{-}$. The diagonal adiabatic correction  $Q_{\nu\nu}(R)$ stays very stable in proportion to $1/R^{2}$, while it varies asymptotically as $1/R^{3}$ for the asymptotic channels that retain a non-interacting value of $l_{\text{eff}}$. Using different poles of the Lennard-Jones potential (i.e., two-body potentials supporting different numbers of bound states) and different characteristic lengths of a two-body Gaussian interaction potential, the Born-Oppenheimer potential curves (dashed line) have been computed in a separate calculation. While the Born-Oppenheimer potentials asymptotically are found to differ depending on the two-body potential utilized, the adiabatic potential curves (solid line) are found to agree and to have a universal coefficient of $1/R^{2}$.

The case of Fig.(\ref{fig:spsJ1mUW}), for two-component fermions, the interaction between different spins is set at $s$-wave unitarity while between same spins it is set at $p$-wave unitarity.  From our calculation,  $Q_{\nu\nu}(R)$ is proportional to $1/R^{2}$ in the $p$-wave unitary channels at large $R$ and varies asymptotically as $1/R^{3}$ in the $s$-wave unitary channels or in the channels having non-interacting asymptotics. In Fig.(\ref{fig:2FpppJ1mUW}), a pair of fermions in the same spin state has an interaction at the $p$-wave unitarity limit, and each pair in different spin states also interacts at the $p$-wave unitarity limit. In this situation, we find that there are doubly-degenerate $p$-wave unitary channels asymptotically, and the diagonal $Q_{\nu\nu}(R)$ matrix is also proportional to $1/R^{2}$ in the $1^{\tt{st}}$ and $2^{\tt{nd}}$ degenerate $p$-wave unitary channels (where the $l_e$ values are corresponding $0$ and $2$), and has $1/R^{3}$ with hyperradius in the NI asymptotic channels.  Fig.(\ref{fig:pppJ1mUW}) considers three equal spin fermions at the $p$-wave unitary limit with $J^{\Pi}=1^{-}$ symmetry. The asymptotic coefficient of $1/R^{2}$ is found to be negative for the Born-Oppenheimer potential in the $p$-wave unitary channel but once the diagonal adiabatic correction $Q_{\nu\nu}(R)$ is included, that coefficient in the lowest 3-body channel is very close to zero and thus there is no Efimov effect for this symmetry.  In this case, the $Q_{\nu\nu}(R)$ has a $\propto 1/R^{2}$ dependence on the hyperradius, whereas it decays faster (as $1/R^{3}$) in all of the NI asymptotic channels. 

\begin{figure}[h]
\subfloat[\label{fig:spsu1}]
{\includegraphics[width=9cm]{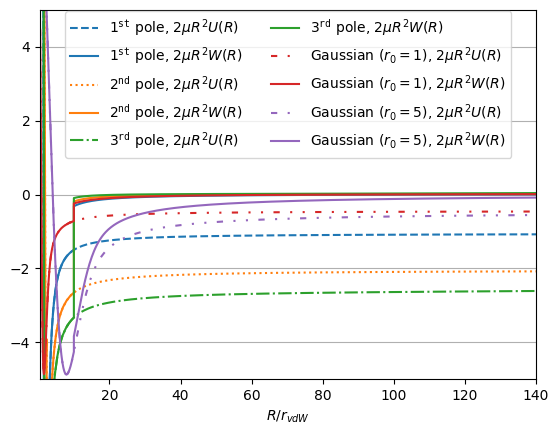}}\centering
\subfloat[\label{fig:spsu2}]
{\includegraphics[width=9cm]{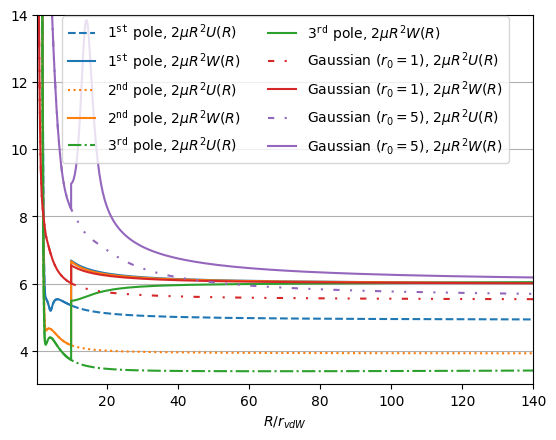}}
\caption{(color online). Comparison of the $p$-wave unitary trimer channels near the first to third $p$-wave resonances for the two-body Lennard-Jones potential and for the first $p$-wave pole of the Gaussian potential with different characteristic lengths. This case is for two component fermionic trimer, where the different fermions in different spin states interact at the $s$-wave unitary limit and where the fermions in the same spin states interact at the $p$-wave unitarity, with overall trimer symmetry $J^{\Pi}=1^{-}$. The dashed curves are the Born-Oppenheimer potentials and the solid curves are the adiabatic potentials that have the diagonal elements of the $Q_{\nu\nu}(R)$ matrix added to the Born-Oppenheimer curves beyond $R=10\, r_{\text{vdW}}$ only. In (a) are shown the first $p$-wave unitary channel with different $p$-wave resonated positions and different two-body potential. In (b) are displayed the second $p$-wave unitary channel for different $p$-wave poles and for different two-body potentials.}
\label{fig:spsu}
\end{figure}

These four figures show the $p$-wave universal behavior  in a number of different cases, i.e. with the fermions in different spin states or with the fermions all in the same spin states, in the symmetry $J^{\Pi}=1^{-}$. Similarly, for other cases from Fig.\ref{fig:sps} to Fig.\ref{fig:ppp}, the $Q_{\nu\nu}(R)$ is also proportional to $1/R^{2}$ in the $p$-wave unitary channels, and they again have an asymptotic dependence proportional to $1/R^{3}$ in the $s$-wave unitary channels and in the channels with non-interacting character at infinity.  Fig.(\ref{fig:spsu}) demonstrates universality in the following sense: using different two-body $p$-wave poles (i.e. with different numbers of two-body bound states) for the Lennard-Jones potential and also for the $1^{\tt{st}}$ $p$-wave pole with different characteristic lengths in a two-body Gaussian potential, we obtain similar {\it adiabatic} potential curves, but different Born-Oppenheimer potentials.

The Gaussian potential can be written as 
\begin{equation}
	v(r) = v_0\exp(-r^2/r_0^2)
\end{equation}
where $r_0$ is the characteristic length, with choices of $r_0=1$ and $r_0=5$ in this case. By adjusting $v_0$, we can get the potential parameters characterizing the desired $1^{st}$ $s$-wave and $p$-wave resonant poles to use in calculations of the three-body system. In this case, a pair of equal-spin fermions interacts at the $p$-wave unitary limit and a pair of fermions in different spin states interacts very close to the $s$-wave unitary limit, for a trimer system with total orbital angular momentum $J^{\Pi}=1^{-}$. Fig.(\ref{fig:spsu1}) and Fig.(\ref{fig:spsu2}) show the first and second $p$-wave unitary channels, respectively. In these figures, the Born-Oppenheimer potential curves exhibit  {\it different coefficients} of $1/(2\mu R^{2})$ asymptotically for the various $p$-wave poles of Lennard-Jones potential or for different  characteristic lengths of the Gaussian potential. On the other hand, in Fig.(\ref{fig:spsu1}), the $p$-wave Efimov phenomenon would initially seem to arise if only the Born-Oppenheimer approximation were utilized for these different two-body potentials or varied $p$-wave poles of the Lennard-Jones potential. However, the adiabatic potential curves would be close together at $R\rightarrow\infty$ (also in Fig.(\ref{fig:spsu2}) case) and the $p$-wave Efimov effect would vanish when adding the adiabatic correction $Q_{\nu\nu}(R)$ diagonal element. Hence, according to our calculation, the $p$-wave unitary channels have universal properties in the three-body system, but no Efimov effect exists for any $p$-wave universal channels with equal mass trimer consisting of one-component or two-components of internal spin. 

\onecolumngrid

%%%%%%%%%%%%%%%%%%%%%%%%%%%%%%%%%%%%%%%%%%%%%%%%%%%%%
%%%%%%%%%%%%%%%%%%%%%%%%%%%%%%%%%%%%%%%%%%%%%%%%%%%%%
\vskip 0.3in

\twocolumngrid

\subsection{Analysis of the $p$-wave unitary channels and the quasi-bound state}
\begin{figure}[h]
\subfloat[\label{fig:spsq1}]
{\includegraphics[width=9cm]{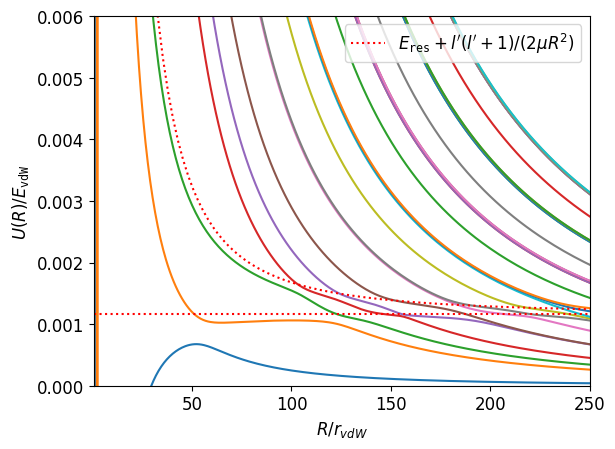}}\centering
\subfloat[\label{fig:spsq2}]
{\includegraphics[width=9cm]{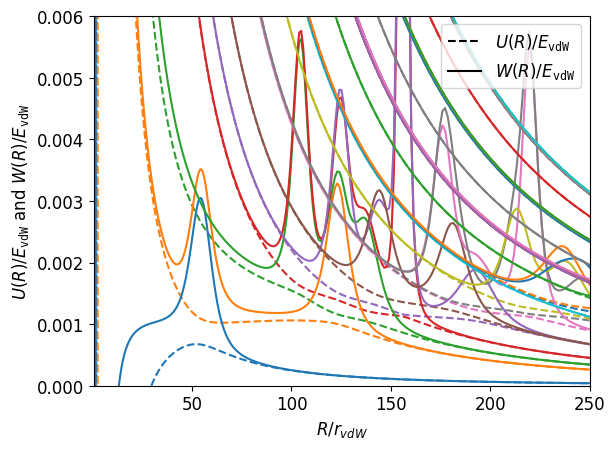}}
\caption{(color online). Shown are Born-Oppenheimer and adiabatic potential curves from channels 1-30, with the two-component fermionic interaction approaching $s$-wave unitarity and the two spin-up fermions having a scattering volume equal to $V_{p}=-1000\, r_{\text{vdW}}^{3}$, for the total trimer symmetry $J^{\Pi}=1^{-}$. In (a), the Born-Oppenheimer potential curves show a series of avoided crossings near the two-body shape resonance $E_{res}=0.0012\, E_{\text{vdW}}$. The two dotted curves show the purely centrifugal curves (that approximate the true diabatic curves) of the form $E_{res}+l'(l'+1)/(2\mu R^{2})$ with $l'=0$ and $2$, respectively. %{\bf Yu-Hsin:  please explain to me what you mean by ``diabatic'' in the preceding sentence.  Do you mean in stead ``purely centrifugal''? To Professor: diabatic curves are represented the avoid-crossing in the Born-Oppenheimer potential.} 
In (b) the Born-Oppenheimer potential curves (dashed curves) and adiabatic potential curves (solid curves) are compared.}
\label{fig:spsq}
\end{figure}

When a quasi-bound state is present, as in Fig.(\ref{fig:spsq}), the Born-Oppenheimer potentials show an infinite set of avoided crossings near the resonance energy. An approximate diabatic potential in this case differs from Eq.(\ref{eq:wasym}), and instead potential takes the following form through those crossings: \cite{suno2003three}
\begin{equation}\label{eq:Eres}
	W(R)=E_{\tt{res}}+l'(l'+1)/(2\mu R^{2})
\end{equation}
where $E_{res}$ corresponds to the two-body shape resonance, i.e. a quasi-bound state energy, and $l'$ is the two-body angular momentum of the third atom with respect to the dimer. Fig.(\ref{fig:spsq}) treats the case where the fermions in different spin states are near an $s$-wave Feshbach resonance and the interaction between two spin-up fermions has $V_{p}= -1000\, r_{\text{vdW}}^{3}$. 

In Fig.(\ref{fig:spsq1}), the Born-Oppenheimer potential curves have an infinite series of avoided crossings because of the two-body $p$-wave resonance energy. The two dotted purely centrifugal curves that approximate the diabatic potentials are described by the above equation; the two-body $p$-wave quasi-bound state energy in this case is $E_{res}=0.0012\, E_{\text{vdW}}$. This $J^{\Pi}=1^{-}$ symmetry allows two values of $l'$, namely $0$ and $2$. When the $p$-wave scattering volume grows to larger negative values and close to divergence, the two-body $p$-wave quasi-bound state energy gets smaller and closer to vanishing. Finally, there are no avoided crossings in the adiabatic potential curves $E_{\tt{res}} \rightarrow 0$, and the second term of the right-hand side of Eq.(\ref{eq:Eres}) remains applicable. Therefore, there are two $p$-wave unitary channels in this case because the symmetry allows two $l'$ values. The two $p$-wave unitary channels can be formed as 
\begin{equation}
	\frac{0}{2\mu R^{2}} \;, \; \frac{6}{2\mu R^{2}}
\end{equation}
for $l'=0$ and $2$, respectively. From this explanation, the $p$-wave unitary channels can be determined with the $l'$ value and the coefficient of $1/(2\mu R^{2})$ can be represented as $l'(l'+1)$. The allowed values of $l'$ can be determined by considering the total orbital angular momentum $J^{\Pi}$. According to our numerical results, all of the $p$-wave unitary channel potentials are extremely close to the representation of Eq.(\ref{eq:Eres}) with various symmetries. This interpretation can be used to interpret why there is  only one $p$-wave unitary channel of the form  $2/(2\mu R^{2})$ for the symmetries $J^{\Pi}=0^{+}$ and $1^{+}$, and why there is also only one $p$-wave unitary channel in the symmetry $J^{\Pi}=2^{-}$ where the coefficient of $1/(2\mu R^{2})$ is close to $6$. This $p$-wave universal behavior would not be modified by the spin state of the Fermi gas, because either three spin-up fermions at the $p$-wave unitary limit or two-component fermionic trimer at $p$-wave unitarity can also be interpreted by this explanation that accounts for the $p$-wave universal behavior in various symmetries. 

In Fig.(\ref{fig:spsq2}), the Born-Oppenheimer potential curves $U_{\nu}(R)$ have several avoided crossings, which cause corresponding peaks in the adiabatic potential curves $W_{\nu}(R)$, as is always the case at an avoided crossing. The first and third channels (blue and green curves) are greatly affected by the diagonal $Q_{\nu\nu}(R)$ elements until the hyperradius gets beyond the avoided crossings. As previously mentioned, the $Q_{\nu\nu}(R)$ diagonal adiabatic correction is proportional to $R^{-2}$ in the $p$-wave unitary channel. Hence there are big differences between the Born-Oppenheimer and adiabatic potential curves before and through the avoided crossings.  In Fig.(\ref{fig:spsq2}), the $p$-wave interaction is not very strong, there is no $p$-wave unitary channel out to infinity since $|V_p|$ is finite and not too large. Beyond the avoided crossing, the $1^{\tt{st}}$ and $3^{\tt{rd}}$ channel adiabatic potential curves merge with Born-Oppenheimer potential curves since the $Q_{\nu\nu}(R)$ matrix elements are proportional to $R^{-3}$ and vanish faster than $R^{-2}$ asymptotically. Beyond the avoided crossing, the $p$-wave unitary channel would change to an $s$-wave unitary channel or NI asymptotic channels because the interaction between two spin-up fermions is not very strong and cannot form a stable $p$-wave unitary channel. If the $p$-wave scattering volume goes to the unitary limit, there are no avoided crossings (nor associated peaks) in the Born-Oppenheimer (adiabatic) potential curves. The $p$-wave unitary channel can be clearly identified in the adiabatic potential curves. The fermionic trimer having three equal spin states and also the trimer of two-component fermions at the $p$-wave unitary limit can both be explained similarly.

\onecolumngrid

%%%%%%%%%%%%%%%%%%%%%%%%%%%%%%%%%%%%%%%%%%%%%%%%%%%%%
%%%%%%%%%%%%%%%%%%%%%%%%%%%%%%%%%%%%%%%%%%%%%%%%%%%%%
\vskip 0.3in
\twocolumngrid

\subsection{Further details about the $p$-wave universality}
Table.(\ref{table:sps}) summarizes our findings for trimers composed of two-component fermions at the $s$-wave unitary limit and with the two spin-up fermions interacting at the $p$-wave unitarity, for  various symmetries $J^{\Pi}$. Each $p$-wave unitary channel occurs as the lowest adiabatic potential curve for that symmetry and $l_e$ [in the  coefficient $l_e(l_e+1)$ of $1/(2\mu R^{2})$] approaches an integer value ($0$, $1$ or $2$). In particular, the $s$-wave universal behavior can also be found for the symmetries $J^{\Pi}=0^{+}$ and $1^{-}$ with the coefficients of $1/(2\mu R^{2})$ very close to the results determined previously by Werner and Castin and by Blume {\it et al.}. In the $J^{\Pi}=1^{+}$ and $2^{-}$ systems, the $s$-wave unitary channels do not occur, owing to symmetry. The $p$-wave unitary channel appears in the lowest potential curve and its $l_{e}$ values are close to an integer. 

Table.(\ref{table:psp}), summarizes our results for 3 fermions in 2 different spin states at the $p$-wave unitary limit and where the interaction between two equal-spin fermions are comparatively weak, with a $p$-wave scattering volume equal to $V_{p}=-2\, r_{\text{vdW}}^{3}$. The $p$-wave unitary channels are also found in the lowest continuum adiabatic potential curve for various other symmetries, and the coefficients $l_e(l_e+1)$ of $1/(2\mu R^{2})$ are close to an integer value of $l_e$ which can be interpreted through Eq.(\ref{eq:Eres}). Interestingly, the NI asymptotic channels are not changed when any pair of fermions interacts at the $p$-wave unitary limit. For example, for any non-interacting three-fermion system, the adiabatic potential curves have only the half-odd-integer values of $l_{e}$ (See Table.(\ref{table:psp})).  However, when the $V_{p}\rightarrow \infty$, the $p$-wave unitary channel emerges at the lowest three-body continuum channel. The similar situation can be found in the Table.(\ref{table:sps}) while the $V_{p}$ is very small, the adiabatic potential curves shows the $s$-wave unitary channels (presumably with an irrational value of $l_{e}$) and non-interacting asymptotic channels (with half-odd-integer values of $l_{e}$). Nevertheless, while the $V_{p}$ is tuned to infinity, a $p$-wave unitary adiabatic channel potential emerges always for the lowest continuum 3-body channel, and in some cases for a few of the higher channels. 

In Table.(\ref{table:2Fppp}), the interaction between two different-spin fermions and that between two equal-spin fermions are both set at the $p$-wave unitary limit for various symmetries $J^{\Pi}$. For this case, doubly-degenerate $p$-wave unitary channels emerge and again, their asymptotic $l_e$ values are close to integer values. This case can be seen as a combination of the preceding two cases; the degenerate $l_e$ values can also be interpreted by using the Eq.(\ref{eq:Eres}) for different symmetries. In Table.(\ref{table:ppp}) we show that for three equal spin fermions at the $p$-wave unitary limit, the $p$-wave unitary channels can be interpreted using Eq.(\ref{eq:Eres}), and in the coefficient of $1/(2\mu R^{2})$, $l_e$ is again found to be near an integer value. In this case, the $p$-wave universal behavior has a pattern similar to the  cases discussed above, where the $p$-wave unitary channel would reliably occur for the lowest adiabatic potential curve, but does not change the coefficient of $R^{-2}$ in the channel that has a non-interacting value of $l_e$ asymptotically.

%%%%%%%%%%%%%%%%%%%%%%%%%%%%%%%%%%%%%%%%%%%%%%%%%%%%
\begin{table}[h]
    \centering
    \subfloat[Subtable 1 list of tables text][$\uparrow\downarrow$ at $s$-wave unitary and $\uparrow\uparrow$ at $p$-wave unitary limit]{
    \resizebox{\columnwidth}{!}{
    \begin{tabular}{c|cccccccccc}
        \hline\hline
        \footnotesize{$J^{\Pi}$} & & & & & $l_{e}$ & & & & & \\
        \hline
        $0^{+}$   & \textbf{1.000} & 1.666 & 4.628 & 6.615 & 7.500 & 8.332 & 9.500 & 10.563 & 11.500 &\\
        \hline
        $1^{+}$   & \textbf{1.000} & 3.500 & 5.500 & 7.500 & 7.500 & 9.500 & 9.500 & 11.500 & 11.500 &\\
        \hline
        $1^{-}$   & \textbf{0.021} & 1.272 & \textbf{2.000} & 3.858 & 4.500 & 5.216 & 6.500 & 6.500  & 7.5553 &\\
        \hline
        $2^{-}$   & \textbf{2.000} & 4.500 & 6.500 & 6.500 & 8.500 & 8.500 & 8.500 & 10.500 & 10.500 &\\
        \hline\hline
    \end{tabular}}
    \label{table:sps}
    }
    \qquad
    \subfloat[Subtable 2 list of tables text][$\uparrow\downarrow$ at $p$-wave unitary limit]{
    \resizebox{\columnwidth}{!}{
    \begin{tabular}{c|cccccccccc}
        \hline\hline
        \footnotesize{$J^{\Pi}$} & & & & & $l_{e}$ & & & & & \\
        \hline
        $0^{+}$   & \textbf{1.004} & 3.500 & 5.500 & 7.500 & 7.500 & 9.500 & 9.500 & 11.500 & 11.500 &\\
        \hline
        $1^{+}$   & \textbf{1.000} & 3.500 & 5.500 & 7.500 & 7.500 & 9.500 & 9.500 & 11.500 & 11.500 &\\
        \hline
        $1^{-}$   & \textbf{0.002} & \textbf{2.001} & 2.500 & 4.500 & 4.500 & 6.500 & 6.500 & 6.500  & 8.500 &\\
        \hline
        $2^{-}$   & \textbf{2.000} & 4.500 & 6.500 & 6.500 & 8.500 & 8.500 & 8.500 & 10.500 & 10.500 &\\
        \hline\hline
    \end{tabular}}
    \label{table:psp}
    }
    \qquad
    \subfloat[Subtable 3 list of tables text][$\downarrow\uparrow\uparrow$ all at $p$-wave unitary limit]{
    \resizebox{\columnwidth}{!}{
    \begin{tabular}{c|cccccccccc}
        \hline\hline
        \footnotesize{$J^{\Pi}$} & & & & & $l_{e}$ & & & & & \\
        \hline
        $0^{+}$   & \textbf{1.000} & \textbf{1.000} &3.500 & 5.500 & 7.500 & 7.500 & 9.500 & 9.500 & 11.500 &\\
        \hline
        $1^{+}$   & \textbf{1.000} & \textbf{1.000} & 3.500 & 5.500 & 7.500 & 7.500 & 9.500 & 9.500 & 11.500 &\\
        \hline
        $1^{-}$   & \textbf{0.008} & \textbf{0.015} & \textbf{2.011} & \textbf{2.013}  & 2.500 & 4.500 & 4.500 & 6.500  & 6.500 &\\
        \hline
        $2^{-}$   & \textbf{2.002} & \textbf{2.002} & 4.500 & 6.500 & 6.500 & 8.500 & 8.500 & 8.500 & 10.500 &\\
        \hline\hline
    \end{tabular}}
    \label{table:2Fppp}
    }
    \qquad
    \subfloat[Subtable 4 list of tables text][$\uparrow\uparrow\uparrow$ at $p$-wave unitary limit]{
    \resizebox{\columnwidth}{!}{
    \begin{tabular}{c|cccccccccc}
        \hline\hline
        \footnotesize{$J^{\Pi}$} & & & & & $l_{e}$ & & & & & \\
        \hline
        $0^{+}$   & \textbf{1.000} & 7.500 &11.500 & 13.500 & 15.500 & 17.500 & 19.500 & 19.500 & 21.500 &\\
        \hline
        $1^{+}$   & \textbf{1.000} & 3.500 & 7.500 & 9.500 & 11.500 & 13.500 & 15.500 & 15.500 & 17.500 &\\
        \hline
        $1^{-}$   & \textbf{0.011} & \textbf{1.998} & 4.500 & 6.500 & 8.500 & 10.500 & 10.500 & 12.500  & 12.500 &\\
        \hline
        $2^{-}$   & \textbf{2.000} & 6.500 & 8.500 & 10.500 & 12.500 & 12.500 & 14.500 & 14.500 & 16.500 &\\
        \hline\hline
    \end{tabular}}
    \label{table:ppp}
    }
    \caption{Comparison of the $l_{e}$ values (See Eq.(\ref{eq:wasym})) from the first to ninth channels with the different symmetries $J^{\Pi}$, $l_{e}$ is the effective angular momentum which controls the barrier of the three free asymptotic particles in the large hyperradius limit ($R\rightarrow\infty$). The bold font approximately integer value of $l_e$ and the half-integer value of $l_e$  represent the $p$-wave unitary channel and the non-interacting (NI) asymptotic channels, respectively. The other values (which may approximate irrational values of $l_e$, as arise in the Efimov effect) represent $s$-wave unitary channels. In (a), the different-spin fermion interaction is set close to the $s$-wave unitary limit and the interaction between two equal-spin fermions is set close to the $p$-wave unitary limit. In (b), the different-spin fermions interact at the $p$-wave unitary limit for the scattering volume and the two equal-spin fermions have a comparatively weak $p$-wave interaction equal to $V_{p}=-2\,r_{\tt{vdW}}^{3}$. In (c), the interaction between both the different spin and the two same-spin fermions interact at the $p$-wave unitary limit. In (d), the three spin-polarized fermions all interact at the $p$-wave unitary limit.}
\end{table}
%%%%%%%%%%%%%%%%%%%%%%%%%%%%%%%%%%%%%%%%%%%%%%%%%%%%

\section{Conclusions}
This study of the $p$-wave universality implications for various symmetries predicts that either  one or two $p$-wave unitary channels emerge at unitarity, in multiple scenarios where the two-body $p$-wave scattering volume diverges for either a single-component or a two-component fermionic trimer.  The diagonal adiabatic correction matrix element $Q_{\nu\nu}(R)$ plays a key role in determiming the most physically relevant {\it adiabatic} potential curves; in the end the $p$-wave Efimov effect (which would require a complex value of $l_e$) does not occur for any of the symmetries studied:  namely  $J^{\Pi}=0^{+}$, $1^{+}$,$1^{-}$ and $2^{-}$. The $p$-wave universality in the three-body potential curves can be used Eq.(\ref{eq:wasym}) or Eq.(\ref{eq:Eres}) to interpret these findings. Our numerical results obtain a stable, constant coefficient of $1/(2\mu R^{2})$ characterizing the asymptotic channels.  Those values of $l_e$ are crucial, for instance, in determining threshold law exponents for inelastic processes such as three-body recombination. The reduction at $s$-wave or $p$-wave unitarity of the asymptotic coefficients of $1/(2\mu R^{2})$ found in this study can be viewed as examples of the workings of Efimov physics; but these cases considered here with fermionic equal-mass particles never produces a  coefficient reduction to a negative value, and hence there is no Efimov effect that would produce an infinite number of bound states at unitarity.  
%See \cite{JungenAtabek1977}.
%\vskip -6in
\section{Acknowledgements}
This work was supported in part by NSF grant No. PHY-1912350.
\newpage
%%%%%%%%%%%%%%%%%%%%%%%%%%%%%%%%%%%%%%%%%%%%%%%%%%%%
%\begin{table}[h]
%\caption{Comparison of results with those obtained by Castin\cite{werner2006unitary} and %Blume\cite{blume2007universal} with different total angular momentum $J^{\Pi}=0^{+}$ and %$J^{\Pi}=1^{-}$ in the 
%three-fermion system ($\downarrow\uparrow\uparrow$). $E_{00}$ is the ground state of the fermi gas %in the harmonic trap, $E_{10}$ and $E_{20}$ are the first and second excited energy levels, %respectively. }
%\footnotesize
%\begin{tabular}{c|ccc|ccc}
%\hline\hline
%     &.    & $J^{\Pi}=0^{+}$&           &     & $J^{\Pi}=1^{-}$& \\
%\hline 
% $E$ & Our & D. Blume & Y. Castin & Our & D. Blume & Y. Castin \\
%\hline
% $E_{00}$/($\hbar \omega$)  & 4.666  & 4.682 & 4.666 & 4.272 & 4.275 & 4.272   \\
% $E_{10}$/($\hbar \omega$)  & 7.628  & 7.637 & 7.627 & 6.861 & 6.868 & 6.858   \\
% $E_{20}$/($\hbar \omega$)  & 9.615  & 9.628 & 9.614 & 8.215 & 8.229 & 8.216   \\
%\hline\hline
%\end{tabular}
%\label{table:compare}
%\end{table}
%\normalsize
%%%%%%%%%%%%%%%%%%%%%%%%%%%%%%%%%%%%%%%%%%%%%%%%%%%%

%\bibliographystyle{ieeetr} 
\bibliography{references} 

\end{document}